\documentclass[]{article}
\usepackage{lmodern}
\usepackage{amssymb,amsmath}
\usepackage{ifxetex,ifluatex}
\usepackage{fixltx2e} 
\ifnum 0\ifxetex 1\fi\ifluatex 1\fi=0 
  \usepackage[T1]{fontenc}
  \usepackage[utf8]{inputenc}
\else 
  \ifxetex
    \usepackage{mathspec}
  \else
    \usepackage{fontspec}
  \fi
  \defaultfontfeatures{Ligatures=TeX,Scale=MatchLowercase}
\fi
\IfFileExists{upquote.sty}{\usepackage{upquote}}{}
\IfFileExists{microtype.sty}{%
\usepackage{microtype}
\UseMicrotypeSet[protrusion]{basicmath} 
}{}
\usepackage[margin=1in]{geometry}
\usepackage{hyperref}
\hypersetup{unicode=true,
            pdftitle={Tools for analyzing R code the tidy way},
            pdfauthor={Lucy D'Agostino McGowan; Sean Kross; Jeffrey Leek},
            pdfborder={0 0 0},
            breaklinks=true}
\usepackage{color}
\usepackage{fancyvrb}

\DefineVerbatimEnvironment{Highlighting}{Verbatim}{commandchars=\\\{\}}
\usepackage{framed}
\definecolor{shadecolor}{RGB}{248,248,248}
\newenvironment{Shaded}{\begin{snugshade}}{\end{snugshade}}

\newcommand{\DataTypeTok}[1]{\textcolor[rgb]{0.13,0.29,0.53}{#1}}
\newcommand{\DecValTok}[1]{\textcolor[rgb]{0.00,0.00,0.81}{#1}}

\newcommand{\KeywordTok}[1]{\textcolor[rgb]{0.13,0.29,0.53}{\textbf{#1}}}
\newcommand{\NormalTok}[1]{#1}
\newcommand{\OperatorTok}[1]{\textcolor[rgb]{0.81,0.36,0.00}{\textbf{#1}}}
\newcommand{\OtherTok}[1]{\textcolor[rgb]{0.56,0.35,0.01}{#1}}

\newcommand{\StringTok}[1]{\textcolor[rgb]{0.31,0.60,0.02}{#1}}

\usepackage{graphicx,grffile}
\makeatletter
\def\maxwidth{\ifdim\Gin@nat@width>\linewidth\linewidth\else\Gin@nat@width\fi}
\def\maxheight{\ifdim\Gin@nat@height>\textheight\textheight\else\Gin@nat@height\fi}
\makeatother
\setkeys{Gin}{width=\maxwidth,height=\maxheight,keepaspectratio}
\IfFileExists{parskip.sty}{%
\usepackage{parskip}
}{
\setlength{\parindent}{0pt}
\setlength{\parskip}{6pt plus 2pt minus 1pt}
}
\providecommand{\tightlist}{%
  \setlength{\itemsep}{0pt}\setlength{\parskip}{0pt}}
\ifx\paragraph\undefined\else
\let\oldparagraph\paragraph
\renewcommand{\paragraph}[1]{\oldparagraph{#1}\mbox{}}
\fi
\ifx\subparagraph\undefined\else
\let\oldsubparagraph\subparagraph
\renewcommand{\subparagraph}[1]{\oldsubparagraph{#1}\mbox{}}
\fi

\let\rmarkdownfootnote\footnote%
\def\footnote{\protect\rmarkdownfootnote}

\usepackage{titling}

  \title{Tools for analyzing R code the tidy way}
    \pretitle{\vspace{\droptitle}\centering\huge}
  \posttitle{\par}
    \author{Lucy D'Agostino McGowan \\ Sean Kross \\ Jeffrey Leek}
    \preauthor{\centering\large\emph}
  \postauthor{\par}
    \date{}
    \predate{}\postdate{}
  
\usepackage{booktabs}

\begin{document}
\maketitle
\begin{abstract}
With the current emphasis on reproducibility and replicability, there is
an increasing need to examine how data analyses are conducted. In order
to analyze the between researcher variability in data analysis choices
as well as the aspects within the data analysis pipeline that contribute
to the variability in results, we have created two R packages:
\textbf{matahari} and \textbf{tidycode}. These packages build on methods
created for natural language processing; rather than allowing for the
processing of natural language, we focus on R code as the substrate of
interest. The \textbf{matahari} package facilitates the logging of
everything that is typed in the R console or in an R script in a tidy
data frame. The \textbf{tidycode} package contains tools to allow for
analyzing R calls in a tidy manner. We demonstrate the utility of these
packages as well as walk through two examples.
\end{abstract}

\hypertarget{introduction}{%
\subsection{Introduction}\label{introduction}}

With the current emphasis on reproducibility and replicability, there is
an increasing need to examine how data analyses are conducted (Goecks et
al. 2010; Peng 2011; McNutt 2014; Miguel et al. 2014; Ioannidis et al.
2014; Richard 2014; Leek and Peng 2015; Nosek et al. 2015; Sidi and
Harel 2018). In order to accurately replicate a result, the exact
methods used for data analysis need to be recorded, including the
specific analytic steps taken as well as the software utilized
(Waltemath and Wolkenhauer 2016). Studies across multiple disciplines
have examined the global set of possible data analyses that can
conducted on a specific data set (Silberzhan et al. 2018). While we are
able to define this global set, very little is known about the actual
variation that exists between researchers. For example, it is possible
that the true range of data analysis choices is realistically a much
more narrow set than the global sets that are presented. There is a
breadth of excellent research and experiments examining how people read
visual information (Majumder, Hofmann, and Cook 2013; Loy, Hofmann, and
Cook 2017; Wickham, Cook, and Hofmann 2015; Buja et al. 2009; Loy,
Follett, and Hofmann 2016), for example the Experiments on Visual
Inference detailed here:
(\url{http://mamajumder.github.io/html/experiments.html}), but not how
they actually make analysis choices, specifically analysis \emph{coding}
choices. In addition to not knowing about the ``data analysis choice''
variability between researchers, we also don't know which portions of
the data analysis pipeline result in the most variability in the
ultimate research result. We seek to build tools to analyze these two
aspects of data analysis:

\begin{enumerate}
\def\labelenumi{\arabic{enumi}.}
\tightlist
\item
  The between researcher variability in data analysis choices
\item
  The aspects within the data analysis pipeline that contribute to the
  variability in results
\end{enumerate}

Specifically, we have designed a framework to conduct such analyses and
created two R packages that allow for the study of data analysis code
conducted in R. In addition to answering these crucial questions for
broad research fields, we see these tools having additional concrete use
cases. These tools will facilitate data science and statistics pedagogy,
allowing researchers and instructors to investigate how students are
conducting data analyses in the classroom. Alternatively, a researcher
could use these tools to examine how collaborators have conducted a data
analysis. Finally, these tools could be used in a meta-manner to explore
how current software and tools in R are being utilized.

\hypertarget{tidy-principles}{%
\subsubsection{Tidy principles}\label{tidy-principles}}

We specifically employ \emph{tidy} principles in our proposed packages.
\emph{Tidy} refers to an implementation strategy propagated by Hadley
Wickham and implemented by the Tidyverse team at RStudio (Wickham and
Grolemund 2016). Here, by \emph{tidy} we mean our packages adhere to the
following principles:

\begin{enumerate}
\def\labelenumi{\arabic{enumi}.}
\tightlist
\item
  Our functions follow the principles outlined in \emph{R packages}
  (Wickham 2015) as well as the
  \href{http://style.tidyverse.org}{tidyverse style guide} (Wickham
  2019).
\item
  Our output data sets are tidy, as in:
\end{enumerate}

\begin{itemize}
\tightlist
\item
  Each variable has its own column.
\item
  Each observation has its own row.
\item
  Each value has its own cell.
\end{itemize}

By implementing these tidy principles, and thus outputting tidy data
frames, we allow for data manipulation and analysis to be conducted
using a specific set of tools, such as those included in the
\textbf{tidyverse} meta package (Wickham 2017).

Ultimately, we create a mechanism to utilize methods created for natural
language processing; here the substrate is \emph{code} rather than
natural language. We model our tools to emulate the \textbf{tidytext}
package (Silge and Robinson 2016, 2017); instead of analyzing tokens of
text, we are analyzing tokens of code.

We present two packages, \textbf{matahari}, a package for logging
everything that is typed in the R console or in an R script, and
\textbf{tidycode}, a package with tools to allow for analyzing R calls
in a tidy manner. In this paper, we first explain how these packages
work. We then demonstrate two examples, one that analyzes data collected
from an online experiment, and one that analyzes ``old'' data via
previously created R scripts.

\hypertarget{methods}{%
\subsection{Methods}\label{methods}}

We have created two R packages, \textbf{matahari} and \textbf{tidycode}.
The former is a way to log R code, the latter allows the user to analyze
R calls on the function-level in a tidy manner. Figure 1 is a flowchart
of the process described in more detail below. This flowchart is adapted
from Figure 2.1 in \emph{Text Mining with R: A Tidy Approach} (Silge and
Robinson 2017).

\begin{figure}
\includegraphics[width=6.5in]{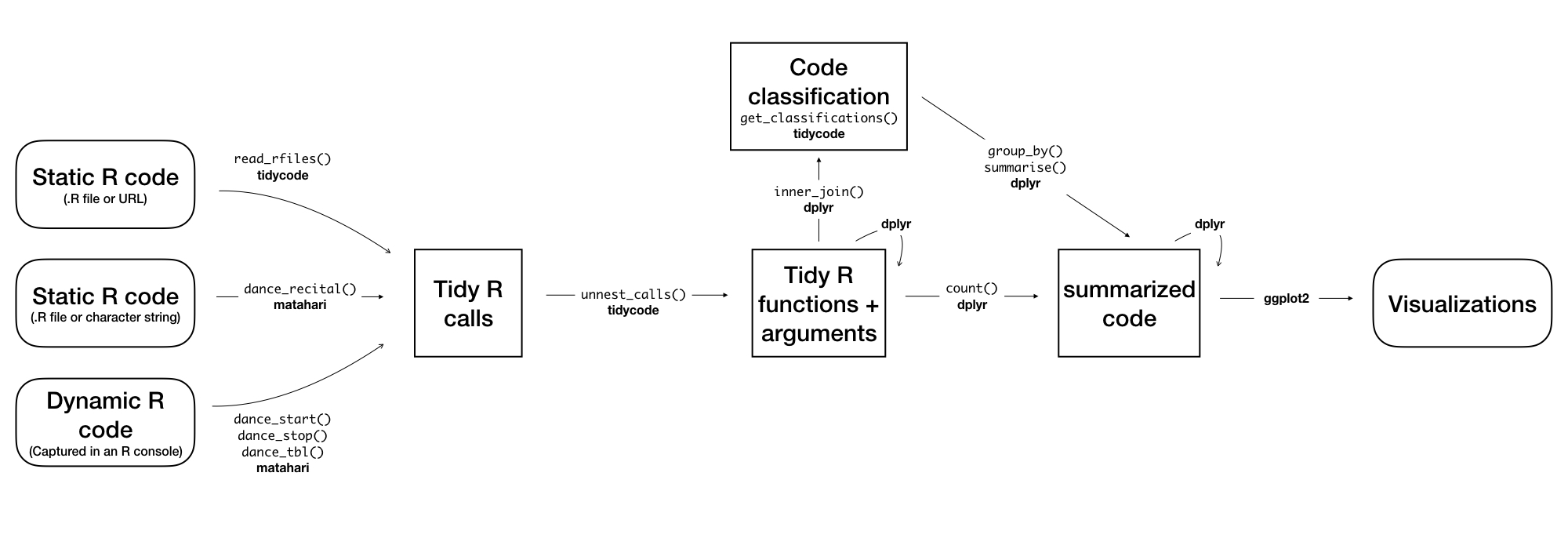} \caption{A flowchart of a typical analysis that uses \textbf{matahari} and \textbf{tidycode} to analyze and classify R code.}\label{fig:fig1}
\end{figure}

We demonstrate how to create these tidy data frames of R code and then
emulate the data analysis workflow similar to that put forth in the tidy
text literature.

\hypertarget{terminology}{%
\subsubsection{Terminology}\label{terminology}}

In this paper, we refer to R ``expressions'' or ``calls'' as well as R
``functions'' and ``arguments''. An R call is a combination of an R
function with arguments. For example, the following is an R call
(Example 1).

\begin{Shaded}
\begin{Highlighting}[]
\KeywordTok{library}\NormalTok{(tidycode)}
\end{Highlighting}
\end{Shaded}

\noindent \textbf{Example 1.} R call, library

Another example of an R call is the following piped chain of functions
from the \textbf{dplyr} package (Example 2).

\begin{Shaded}
\begin{Highlighting}[]
\NormalTok{starwars }\OperatorTok{%>%}
\StringTok{  }\KeywordTok{select}\NormalTok{(height, mass)}
\end{Highlighting}
\end{Shaded}

\noindent \textbf{Example 2.} Piped R call

Specifically, we know something is a call in R if \texttt{is.call()} is
\texttt{TRUE}.

\begin{Shaded}
\begin{Highlighting}[]
\KeywordTok{quote}\NormalTok{(starwars }\OperatorTok{%>%}
\StringTok{  }\KeywordTok{select}\NormalTok{(height, mass)) }\OperatorTok{%>%}
\StringTok{  }\KeywordTok{is.call}\NormalTok{()}
\end{Highlighting}
\end{Shaded}

\begin{verbatim}
## [1] TRUE
\end{verbatim}

Calls in R are made up of a function or name of a function, and
arguments. For example, the call \texttt{library(tidycode)} from Example
1 is comprised of the function \texttt{library()} and the argument
\texttt{tidycode}. Example 2 is a bit more complicated. The piped code
can be rewritten, as seen in Example 3.

\begin{Shaded}
\begin{Highlighting}[]
\StringTok{`}\DataTypeTok{%>%}\StringTok{`}\NormalTok{(starwars, }\KeywordTok{select}\NormalTok{(height, mass))}
\end{Highlighting}
\end{Shaded}

\noindent \textbf{Example 3.} Rewritten piped R call

From this example, it is easier to see that the function for this R call
is \texttt{\%\textgreater{}\%} with two arguments, \texttt{starwars} and
\texttt{select(height,\ mass)}. Notice that one of these arguments is an
R call itself, \texttt{select(height,\ mass)}.

\hypertarget{matahari}{%
\subsubsection{matahari}\label{matahari}}

\textbf{matahari} is a simple package for logging R code in a tidy
manner. It can be installed from CRAN using the following code.

\begin{Shaded}
\begin{Highlighting}[]
\KeywordTok{install.packages}\NormalTok{(}\StringTok{"matahari"}\NormalTok{)}
\end{Highlighting}
\end{Shaded}

There are three ways to use the matahari package:

\begin{enumerate}
\def\labelenumi{\arabic{enumi}.}
\tightlist
\item
  Record R code as it is typed and output a tidy data frame of the
  contents
\item
  Input a character string of R code and output a tidy data frame of the
  contents
\item
  Input an R file containing R code and output a tidy data frame of the
  contents
\end{enumerate}

In the following sections, we will split these into two categories, tidy
logging from the R console (1) and tidy logging from an R script (2 and
3).

\hypertarget{tidy-logging-from-the-r-console}{%
\paragraph{Tidy logging from the R
console}\label{tidy-logging-from-the-r-console}}

In order to begin logging from the R console, the
\texttt{dance\_start()} function is used. Logging is paused using
\texttt{dance\_stop()} and the log can be viewed using
\texttt{dance\_tbl()}. For example, the following code will result in
the the subsequent tidy data frame.

\begin{Shaded}
\begin{Highlighting}[]
\KeywordTok{library}\NormalTok{(matahari)}
\KeywordTok{dance_start}\NormalTok{()}
\DecValTok{1} \OperatorTok{+}\StringTok{ }\DecValTok{2}
\StringTok{"here is some text"}
\KeywordTok{sum}\NormalTok{(}\DecValTok{1}\OperatorTok{:}\DecValTok{10}\NormalTok{)}
\KeywordTok{dance_stop}\NormalTok{()}
\KeywordTok{dance_tbl}\NormalTok{()}
\end{Highlighting}
\end{Shaded}

\begin{verbatim}
#> # A tibble: 6 x 6
#>   expr        value             path       contents   selection dt                  
#>   <list>      <list>            <list>     <list>     <list>    <dttm>             
#> 1 <languag... <S3: sessionIn... <lgl [1... <lgl [1... <lgl [1]> 2018-09-11 22:22:12
#> 2 <languag... <lgl [1]>         <lgl [1... <lgl [1... <lgl [1]> 2018-09-11 22:22:12
#> 3 <languag... <lgl [1]>         <lgl [1... <lgl [1... <lgl [1]> 2018-09-11 22:22:12
#> 4 <chr [1]>   <lgl [1]>         <lgl [1... <lgl [1... <lgl [1]> 2018-09-11 22:22:12
#> 5 <languag... <lgl [1]>         <lgl [1... <lgl [1... <lgl [1]> 2018-09-11 22:22:12
#> 6 <languag... <S3: sessionIn... <lgl [1... <lgl [1... <lgl [1]> 2018-09-11 22:22:12
\end{verbatim}

\noindent \textbf{Example 4.} Logging R code from the R console using
\textbf{matahari}

The resulting tidy data frame consists of 6 columns: \texttt{expr}, the
R call that was run, \texttt{value}, the value that was output,
\texttt{path}, if the code was run within RStudio, this will be the path
to the file in focus, \texttt{contents}, the file contents of the
RStudio editor tab in focus, \texttt{selection}, the text that is
highlighted in the RStudio editor tab in focus, and \texttt{dt}, the
date and time the expression was run. By default, \texttt{value},
\texttt{path}, \texttt{contents} and \texttt{selection} will not be
logged unless the argument is set to \texttt{TRUE} in the
\texttt{dance\_start()} function. For example, if the analyst wanted the
output data frame to include the values computed, they would input
\texttt{dance\_start(value\ =\ TRUE)}.

In this particular data frame, there are 6 rows. The first and final
rows report the R session information at the time when
\texttt{dance\_start()} was initiated (row 1) and when
\texttt{dance\_stop()} was run (row 6). The second row holds the R call
\texttt{dance\_start()}, the first command run in the R console, was
run; the third row holds \texttt{1\ +\ 2}, the fourth holds
\texttt{here\ is\ some\ text}, and the fifth holds \texttt{sum(1:10)}.

\begin{Shaded}
\begin{Highlighting}[]
\KeywordTok{dance_tbl}\NormalTok{()[[}\StringTok{"expr"}\NormalTok{]]}
\end{Highlighting}
\end{Shaded}

\begin{verbatim}
#> [[1]]
#> sessionInfo()
#> 
#> [[2]]
#> dance_start()
#> 
#> [[3]]
#> 1 + 2
#> 
#> [[4]]
#> [1] "here is some text"
#> 
#> [[5]]
#> sum(1:10)
#> 
#> [[6]]
#> sessionInfo()
\end{verbatim}

These functions work by saving an invisible data frame called
\texttt{.dance} that is referenced by \texttt{dance\_tbl()}. Each time
\texttt{dance\_start()} is subsequently run after
\texttt{dance\_stop()}, new rows of data are added to this data frame.
This invisible data frame exists in a new environment created by the
\textbf{matahari} package. We can remove this data frame by running
\texttt{dance\_remove()}.

This data frame can be manipulated using common R techniques. Below, we
rerun the same code as above, this time saving the values that are
computed in the R console by using the \texttt{value\ =\ TRUE}
parameter.

\begin{Shaded}
\begin{Highlighting}[]
\KeywordTok{dance_start}\NormalTok{(}\DataTypeTok{value =} \OtherTok{TRUE}\NormalTok{)}
\DecValTok{1} \OperatorTok{+}\StringTok{ }\DecValTok{2}
\StringTok{"here is some text"}
\KeywordTok{sum}\NormalTok{(}\DecValTok{1}\OperatorTok{:}\DecValTok{10}\NormalTok{)}
\KeywordTok{dance_stop}\NormalTok{()}
\NormalTok{tbl <-}\StringTok{ }\KeywordTok{dance_tbl}\NormalTok{()}
\end{Highlighting}
\end{Shaded}

As an example of the type of data wrangling that this tidy format allows
for, using \textbf{dplyr} and \textbf{purrr}, we can manipulate this to
only examine expressions that result in numeric values.

\begin{Shaded}
\begin{Highlighting}[]
\KeywordTok{library}\NormalTok{(dplyr)}
\KeywordTok{library}\NormalTok{(purrr)}

\NormalTok{t_numeric <-}\StringTok{ }\NormalTok{tbl }\OperatorTok{%>%}
\StringTok{  }\KeywordTok{mutate}\NormalTok{(}
    \DataTypeTok{numeric_output =} \KeywordTok{map_lgl}\NormalTok{(value, is.numeric)}
\NormalTok{  ) }\OperatorTok{%>%}
\StringTok{  }\KeywordTok{filter}\NormalTok{(numeric_output)}

\NormalTok{t_numeric}
\end{Highlighting}
\end{Shaded}

\begin{verbatim}
#> # A tibble: 3 x 7
#>   expr       value     path      contents  selection dt                  numeric_output
#>   <list>     <list>    <list>    <list>    <list>    <dttm>              <lgl>         
#> 1 <language> <int [1]> <lgl [1]> <lgl [1]> <lgl [1]> 2019-04-29 22:39:05 TRUE          
#> 2 <language> <dbl [1]> <lgl [1]> <lgl [1]> <lgl [1]> 2019-04-29 22:39:05 TRUE          
#> 3 <language> <int [1]> <lgl [1]> <lgl [1]> <lgl [1]> 2019-04-29 22:39:05 TRUE 
\end{verbatim}

Here, three rows are output, since we have filtered to only calls with
numeric output:

\begin{enumerate}
\def\labelenumi{\arabic{enumi}.}
\tightlist
\item
  The \texttt{dance\_start()} call (this defaults to have a numeric
  value of 1)
\item
  The \texttt{1\ +\ 2} call, resulting in a \texttt{value} of \texttt{3}
\item
  The \texttt{sum(1:10)}, resulting in a \texttt{value} of \texttt{55}
\end{enumerate}

\hypertarget{tidy-logging-from-an-r-script}{%
\paragraph{Tidy logging from an R
script}\label{tidy-logging-from-an-r-script}}

In addition to allowing for the logging of everything typed in the R
console, the \textbf{matahari} package also allows for the logging of
pre-created R scripts. This can be done using the
\texttt{dance\_recital()} function, which allows for either a .R file or
a character string of R calls as the input. For example, if we have a
code file called \texttt{sample\_code.R}, we can run
\texttt{dance\_recital("sample\_code.R")} to create a tidy data frame.
Alternatively, we can enter code directly as a string of text, such as
\texttt{dance\_recital("1\ +\ 2")} to create the tidy data frame. Below
illustrates this functionality.

\begin{Shaded}
\begin{Highlighting}[]
\NormalTok{code_file <-}\StringTok{ }\KeywordTok{system.file}\NormalTok{(}\StringTok{"test"}\NormalTok{, }\StringTok{"sample_code.R"}\NormalTok{, }\DataTypeTok{package =} \StringTok{"matahari"}\NormalTok{)}
\KeywordTok{dance_recital}\NormalTok{(code_file)}
\end{Highlighting}
\end{Shaded}

\begin{verbatim}
#> # A tibble: 7 x 6
#>   expr       value     error             output    warnings  messages 
#>   <list>     <list>    <list>            <list>    <list>    <list>   
#> 1 <language> <dbl [1]> <NULL>            <chr [1]> <chr [0]> <chr [0]>
#> 2 <chr [1]>  <chr [1]> <NULL>            <chr [1]> <chr [0]> <chr [0]>
#> 3 <language> <dbl [1]> <NULL>            <chr [1]> <chr [0]> <chr [0]>
#> 4 <language> <NULL>    <S3: simpleError> <NULL>    <NULL>    <NULL>   
#> 5 <language> <chr [1]> <NULL>            <chr [1]> <chr [1]> <chr [0]>
#> 6 <language> <NULL>    <NULL>            <chr [1]> <chr [0]> <chr [1]>
#> 7 <language> <NULL>    <NULL>            <chr [1]> <chr [0]> <chr [0]>
\end{verbatim}

\noindent \textbf{Example 5.} R call, Logging code from a .R file using
\textbf{matahari}

\begin{Shaded}
\begin{Highlighting}[]
\NormalTok{code_string <-}\StringTok{ '}
\StringTok{4 + 4}
\StringTok{"wow!"}
\StringTok{mean(1:10)}
\StringTok{stop("Error!")}
\StringTok{warning("Warning!")}
\StringTok{message("Hello?")}
\StringTok{cat("Welcome!")}
\StringTok{'}
\KeywordTok{dance_recital}\NormalTok{(code_string)}
\end{Highlighting}
\end{Shaded}

\begin{verbatim}
#> # A tibble: 7 x 6
#>   expr       value     error             output    warnings  messages 
#>   <list>     <list>    <list>            <list>    <list>    <list>   
#> 1 <language> <dbl [1]> <NULL>            <chr [1]> <chr [0]> <chr [0]>
#> 2 <chr [1]>  <chr [1]> <NULL>            <chr [1]> <chr [0]> <chr [0]>
#> 3 <language> <dbl [1]> <NULL>            <chr [1]> <chr [0]> <chr [0]>
#> 4 <language> <NULL>    <S3: simpleError> <NULL>    <NULL>    <NULL>   
#> 5 <language> <chr [1]> <NULL>            <chr [1]> <chr [1]> <chr [0]>
#> 6 <language> <NULL>    <NULL>            <chr [1]> <chr [0]> <chr [1]>
#> 7 <language> <NULL>    <NULL>            <chr [1]> <chr [0]> <chr [0]>
\end{verbatim}

\noindent \textbf{Example 6.} Logging code from a character string using
\textbf{matahari}

The resulting tidy data frame from \texttt{dance\_recital()}, as seen in
Examples 5 and 6, is different from that of \texttt{dance\_tbl()}. This
data frame has 6 columns. The first is the same as the
\texttt{dance\_tbl()}, \texttt{expr}, the R calls in the .R script or
string of code. The subsequent columns are, \texttt{value}, the computed
result of the R call, \texttt{error}, which contains the resulting error
object from a poorly formed call, \texttt{output}, the printed output
from an call, \texttt{warnings}, the contents of any warnings that would
be displayed in the console, and \texttt{messages}, the contents of any
generated diagnostic messages. Now that we have a tidy data frame with R
calls obtained either from the R console or from a .R script, we can
analyze them using the \textbf{tidycode} package.

\hypertarget{tidycode}{%
\subsubsection{tidycode}\label{tidycode}}

The goal of \textbf{tidycode} is to allow users to analyze R scripts,
calls, and functions in a tidy way. There are two main tasks that can be
achieved with this package:

\begin{enumerate}
\def\labelenumi{\arabic{enumi}.}
\tightlist
\item
  We can ``tokenize'' R calls
\item
  We can classify the functions run into one of nine potential data
  analysis categories: ``Setup'', ``Exploratory'', ``Data Cleaning'',
  ``Modeling'', ``Evaluation'',``Visualization'', ``Communication'',
  ``Import'', or ``Export''.
\end{enumerate}

The \textbf{tidycode} package can be installed from CRAN in the
following manner.

\begin{Shaded}
\begin{Highlighting}[]
\KeywordTok{install.packages}\NormalTok{(}\StringTok{"tidycode"}\NormalTok{)}
\end{Highlighting}
\end{Shaded}

\begin{Shaded}
\begin{Highlighting}[]
\KeywordTok{library}\NormalTok{(tidycode)}
\end{Highlighting}
\end{Shaded}

We can first create a tidy data frame using the \textbf{matahari}
package. Alternatively, we can use a function in the \textbf{tidycode}
package that wraps the \texttt{dance\_recital()} function called
\texttt{read\_rfiles()}. This function allows you to read in multiple .R
files or links to .R files. There are a few example files included in
the \textbf{tidycode} package. The paths to these files can be accessed
via the \texttt{tidycode\_example()} function. For example, running the
following code will give the file path for the
\texttt{example\_analysis.R} file.

\begin{Shaded}
\begin{Highlighting}[]
\KeywordTok{tidycode_example}\NormalTok{(}\StringTok{"example_analysis.R"}\NormalTok{)}
\end{Highlighting}
\end{Shaded}

\begin{verbatim}
## [1] "/Library/Frameworks/R.framework/Versions/3.5/Resources/library/tidycode/extdata/example_analysis.R"
\end{verbatim}

Running the function without any file specified will supply a vector of
all available file names.

\begin{Shaded}
\begin{Highlighting}[]
\KeywordTok{tidycode_example}\NormalTok{()}
\end{Highlighting}
\end{Shaded}

\begin{verbatim}
## [1] "example_analysis.R" "example_plot.R"
\end{verbatim}

We can use these example files in the \texttt{read\_rfiles()} function.

\begin{Shaded}
\begin{Highlighting}[]
\NormalTok{df <-}\StringTok{ }\KeywordTok{read_rfiles}\NormalTok{(}\KeywordTok{tidycode_example}\NormalTok{(}\KeywordTok{c}\NormalTok{(}\StringTok{"example_analysis.R"}\NormalTok{, }\StringTok{"example_plot.R"}\NormalTok{)))}
\NormalTok{df}
\end{Highlighting}
\end{Shaded}

\begin{verbatim}
## # A tibble: 9 x 3
##   file                                                       expr      line
##   <chr>                                                      <list>   <int>
## 1 /Library/Frameworks/R.framework/Versions/3.5/Resources/li~ <langua~     1
## 2 /Library/Frameworks/R.framework/Versions/3.5/Resources/li~ <langua~     2
## 3 /Library/Frameworks/R.framework/Versions/3.5/Resources/li~ <langua~     3
## 4 /Library/Frameworks/R.framework/Versions/3.5/Resources/li~ <langua~     4
## 5 /Library/Frameworks/R.framework/Versions/3.5/Resources/li~ <langua~     5
## 6 /Library/Frameworks/R.framework/Versions/3.5/Resources/li~ <langua~     6
## 7 /Library/Frameworks/R.framework/Versions/3.5/Resources/li~ <langua~     7
## 8 /Library/Frameworks/R.framework/Versions/3.5/Resources/li~ <langua~     1
## 9 /Library/Frameworks/R.framework/Versions/3.5/Resources/li~ <langua~     2
\end{verbatim}

This will give a tidy data frame with three columns: \texttt{file}, the
path to the file, \texttt{expr} the R call, and \texttt{line} the line
the call was made in the original .R file.

We can then use the \texttt{unnest\_calls()} function to create a data
frame of the calls, splitting each into the individual functions and
arguments. We liken this to the \textbf{tidytext}
\texttt{unnest\_tokens()} function. This function has two parameters,
\texttt{.data}, the data frame that contains the R calls, and
\texttt{input} the name of the column that contains the R calls. In this
case, the data frame is \texttt{m} and the input column is
\texttt{expr}.

\begin{Shaded}
\begin{Highlighting}[]
\NormalTok{u <-}\StringTok{ }\KeywordTok{unnest_calls}\NormalTok{(df, expr)}
\NormalTok{u}
\end{Highlighting}
\end{Shaded}

\begin{verbatim}
## # A tibble: 35 x 4
##    file                                                line func   args    
##    <chr>                                              <int> <chr>  <list>  
##  1 /Library/Frameworks/R.framework/Versions/3.5/Reso~     1 libra~ <list [~
##  2 /Library/Frameworks/R.framework/Versions/3.5/Reso~     2 libra~ <list [~
##  3 /Library/Frameworks/R.framework/Versions/3.5/Reso~     3 <-     <list [~
##  4 /Library/Frameworks/R.framework/Versions/3.5/Reso~     3 %>%    <list [~
##  5 /Library/Frameworks/R.framework/Versions/3.5/Reso~     3 %>%    <list [~
##  6 /Library/Frameworks/R.framework/Versions/3.5/Reso~     3 mutate <list [~
##  7 /Library/Frameworks/R.framework/Versions/3.5/Reso~     3 /      <list [~
##  8 /Library/Frameworks/R.framework/Versions/3.5/Reso~     3 (      <list [~
##  9 /Library/Frameworks/R.framework/Versions/3.5/Reso~     3 ^      <list [~
## 10 /Library/Frameworks/R.framework/Versions/3.5/Reso~     3 (      <list [~
## # ... with 25 more rows
\end{verbatim}

This results is a tidy data frame with two additional columns:
\texttt{func} the name of the function called and \texttt{args} the
arguments of the function called. Because this function takes a data
frame as the first argument, it works nicely with the tidyverse data
manipulation packages. For example, we could get the same data frame as
above by using the following code.

\begin{Shaded}
\begin{Highlighting}[]
\NormalTok{df }\OperatorTok{%>%}
\StringTok{  }\KeywordTok{unnest_calls}\NormalTok{(expr)}
\end{Highlighting}
\end{Shaded}

\begin{verbatim}
## # A tibble: 35 x 4
##    file                                                line func   args    
##    <chr>                                              <int> <chr>  <list>  
##  1 /Library/Frameworks/R.framework/Versions/3.5/Reso~     1 libra~ <list [~
##  2 /Library/Frameworks/R.framework/Versions/3.5/Reso~     2 libra~ <list [~
##  3 /Library/Frameworks/R.framework/Versions/3.5/Reso~     3 <-     <list [~
##  4 /Library/Frameworks/R.framework/Versions/3.5/Reso~     3 %>%    <list [~
##  5 /Library/Frameworks/R.framework/Versions/3.5/Reso~     3 %>%    <list [~
##  6 /Library/Frameworks/R.framework/Versions/3.5/Reso~     3 mutate <list [~
##  7 /Library/Frameworks/R.framework/Versions/3.5/Reso~     3 /      <list [~
##  8 /Library/Frameworks/R.framework/Versions/3.5/Reso~     3 (      <list [~
##  9 /Library/Frameworks/R.framework/Versions/3.5/Reso~     3 ^      <list [~
## 10 /Library/Frameworks/R.framework/Versions/3.5/Reso~     3 (      <list [~
## # ... with 25 more rows
\end{verbatim}

We can further manipulate this, for example we could select just the
\texttt{func} and \texttt{args} columns using \textbf{dplyr}'s
\texttt{select()} function.

\begin{Shaded}
\begin{Highlighting}[]
\NormalTok{df }\OperatorTok{%>%}
\StringTok{  }\KeywordTok{unnest_calls}\NormalTok{(expr) }\OperatorTok{%>%}
\StringTok{  }\KeywordTok{select}\NormalTok{(func, args)}
\end{Highlighting}
\end{Shaded}

\begin{verbatim}
## # A tibble: 35 x 2
##    func    args      
##    <chr>   <list>    
##  1 library <list [1]>
##  2 library <list [1]>
##  3 <-      <list [2]>
##  4 %>%     <list [2]>
##  5 %>%     <list [2]>
##  6 mutate  <list [1]>
##  7 /       <list [2]>
##  8 (       <list [1]>
##  9 ^       <list [2]>
## 10 (       <list [1]>
## # ... with 25 more rows
\end{verbatim}

The \texttt{get\_classifications()} function calls a classification data
frame that we curated that classifies the individual functions into one
of nine categories: setup, exploratory, data cleaning, modeling,
evaluation, visualization, communication, import, or export. This can
also be merged into the data frame.

\begin{Shaded}
\begin{Highlighting}[]
\NormalTok{u }\OperatorTok{%>%}
\StringTok{  }\KeywordTok{inner_join}\NormalTok{(}\KeywordTok{get_classifications}\NormalTok{()) }\OperatorTok{%>%}
\StringTok{  }\KeywordTok{select}\NormalTok{(func, classification, lexicon, score)}
\end{Highlighting}
\end{Shaded}

\begin{verbatim}
## # A tibble: 322 x 4
##    func    classification lexicon       score
##    <chr>   <chr>          <chr>         <dbl>
##  1 library setup          crowdsource 0.687  
##  2 library import         crowdsource 0.213  
##  3 library visualization  crowdsource 0.0339 
##  4 library data cleaning  crowdsource 0.0278 
##  5 library modeling       crowdsource 0.0134 
##  6 library exploratory    crowdsource 0.0128 
##  7 library communication  crowdsource 0.00835
##  8 library evaluation     crowdsource 0.00278
##  9 library export         crowdsource 0.00111
## 10 library setup          leeklab     0.994  
## # ... with 312 more rows
\end{verbatim}

There are two lexicons for classification, \texttt{crowdsource} and
\texttt{leeklab}. The former was created by volunteers who classified R
code using the \href{https://lucy.shinyapps.io/classify}{classify shiny
application}. The latter was curated by \href{https://jtleek.com}{Jeff
Leek's Lab}. To select a particular lexicon, you can specify the
\texttt{lexicon} parameter. For example, the following code will merge
in the \texttt{crowdsource} lexicon only.

\begin{Shaded}
\begin{Highlighting}[]
\NormalTok{u }\OperatorTok{%>%}
\StringTok{  }\KeywordTok{inner_join}\NormalTok{(}\KeywordTok{get_classifications}\NormalTok{(}\StringTok{"crowdsource"}\NormalTok{)) }\OperatorTok{%>%}
\StringTok{  }\KeywordTok{select}\NormalTok{(func, classification, score)}
\end{Highlighting}
\end{Shaded}

\begin{verbatim}
## # A tibble: 271 x 3
##    func    classification   score
##    <chr>   <chr>            <dbl>
##  1 library setup          0.687  
##  2 library import         0.213  
##  3 library visualization  0.0339 
##  4 library data cleaning  0.0278 
##  5 library modeling       0.0134 
##  6 library exploratory    0.0128 
##  7 library communication  0.00835
##  8 library evaluation     0.00278
##  9 library export         0.00111
## 10 library setup          0.687  
## # ... with 261 more rows
\end{verbatim}

It is possible for a function to belong to multiple classes. This will
result in multiple lines (and multiple classifications) for a given
function. By default, these multiple classifications are included along
with the prevalence of each, indicated by the \texttt{score} column. To
merge in only the most prevalent classification, set the
\texttt{include\_duplicates} option to \texttt{FALSE}.

\begin{Shaded}
\begin{Highlighting}[]
\NormalTok{u }\OperatorTok{%>%}
\StringTok{  }\KeywordTok{inner_join}\NormalTok{(}\KeywordTok{get_classifications}\NormalTok{(}\StringTok{"crowdsource"}\NormalTok{, }\DataTypeTok{include_duplicates =} \OtherTok{FALSE}\NormalTok{)) }\OperatorTok{%>%}
\StringTok{  }\KeywordTok{select}\NormalTok{(func, classification)}
\end{Highlighting}
\end{Shaded}

\begin{verbatim}
## # A tibble: 33 x 2
##    func    classification
##    <chr>   <chr>         
##  1 library setup         
##  2 library setup         
##  3 <-      data cleaning 
##  4 %>%     data cleaning 
##  5 %>%     data cleaning 
##  6 mutate  data cleaning 
##  7 /       data cleaning 
##  8 (       data cleaning 
##  9 ^       modeling      
## 10 (       data cleaning 
## # ... with 23 more rows
\end{verbatim}

In text analysis, there is the concept of ``stopwords''. These are often
small common filler words you want to remove before completing an
analysis, such as ``a'' or ``the''. In a tidy \emph{code} analysis, we
can use a similar concept to remove some functions. For example we may
want to remove the assignment operator, \texttt{\textless{}-}, before
completing an analysis. We have compiled a list of common stop functions
in the \texttt{get\_stopfuncs()} function to anti join from the data
frame.

\begin{Shaded}
\begin{Highlighting}[]
\NormalTok{u }\OperatorTok{%>%}
\StringTok{  }\KeywordTok{inner_join}\NormalTok{(}\KeywordTok{get_classifications}\NormalTok{(}\StringTok{"crowdsource"}\NormalTok{, }\DataTypeTok{include_duplicates =} \OtherTok{FALSE}\NormalTok{)) }\OperatorTok{%>%}
\StringTok{  }\KeywordTok{anti_join}\NormalTok{(}\KeywordTok{get_stopfuncs}\NormalTok{()) }\OperatorTok{%>%}
\StringTok{  }\KeywordTok{select}\NormalTok{(func, classification)}
\end{Highlighting}
\end{Shaded}

\begin{verbatim}
## # A tibble: 15 x 2
##    func       classification
##    <chr>      <chr>         
##  1 library    setup         
##  2 library    setup         
##  3 mutate     data cleaning 
##  4 select     data cleaning 
##  5 options    setup         
##  6 summary    exploratory   
##  7 plot       visualization 
##  8 library    setup         
##  9 select     data cleaning 
## 10 filter     data cleaning 
## 11 is.na      data cleaning 
## 12 is.na      data cleaning 
## 13 ggplot     visualization 
## 14 aes        visualization 
## 15 geom_point visualization
\end{verbatim}

\hypertarget{examples}{%
\subsection{Examples}\label{examples}}

\hypertarget{online-experiment-p-hack-athon}{%
\subsubsection{Online experiment:
P-hack-athon}\label{online-experiment-p-hack-athon}}

This first example demonstrates how to use the \textbf{matahari} and
\textbf{tidycode} packages to analyze data from a prospective study,
using the ``recording'' capabilities of the \textbf{matahari} package to
capture the code as participants run it. Recently, we launched a
``p-hack-athon'' where we encouraged users to analyze a dataset with the
goal of producing the smallest p-value (IRB \# IRB00008885, Not Human
Subjects Research Classification, Johns Hopkins Bloomberg School of
Public Health IRB). We captured the code the participants ran using the
\texttt{dance\_start()} and \texttt{dance\_stop()} functions from the
\textbf{matahari} package. This resulted in a tidy data frame of R calls
for each participant. We use the \textbf{tidycode} package to analyze
these matahari data frames.

\hypertarget{setup}{%
\paragraph{Setup}\label{setup}}

\begin{Shaded}
\begin{Highlighting}[]
\KeywordTok{library}\NormalTok{(tidyverse)}
\KeywordTok{library}\NormalTok{(tidycode)}

\NormalTok{## load the dataset, called df}
\KeywordTok{load}\NormalTok{(}\StringTok{"data/df_phackathon.Rda"}\NormalTok{)}
\end{Highlighting}
\end{Shaded}

The data from the ``p-hack-a-thon'' is saved as a data frame called
\texttt{df}. We have bound the \texttt{expr} column from the
\textbf{matahari} data frame for each participant. Using the
\texttt{unnest\_calls()} function, we unnest each of these R calls into
a function and it's arguments.

\begin{Shaded}
\begin{Highlighting}[]
\NormalTok{tbl <-}\StringTok{ }\NormalTok{df }\OperatorTok{%>%}
\StringTok{  }\KeywordTok{unnest_calls}\NormalTok{(expr)}
\end{Highlighting}
\end{Shaded}

We can then remove the ``stop functions'' by doing an anti join with the
\texttt{get\_stopfuncs()} function and merge in the crowd-sourced
classifications with the \texttt{get\_classifications()} function.

\begin{Shaded}
\begin{Highlighting}[]
\NormalTok{tbl <-}\StringTok{ }\NormalTok{tbl }\OperatorTok{%>%}
\StringTok{  }\KeywordTok{anti_join}\NormalTok{(}\KeywordTok{get_stopfuncs}\NormalTok{()) }\OperatorTok{%>%}
\StringTok{  }\KeywordTok{inner_join}\NormalTok{(}\KeywordTok{get_classifications}\NormalTok{(}\StringTok{"crowdsource"}\NormalTok{, }\DataTypeTok{include_duplicates =} \OtherTok{FALSE}\NormalTok{))}
\end{Highlighting}
\end{Shaded}

\hypertarget{classifications}{%
\paragraph{Classifications}\label{classifications}}

We can use common data manipulation functions from \textbf{dplyr}. For
example, on average, ``data cleaning'' functions made up 39.6\% of the
functions run by participants (Table 1).

\begin{Shaded}
\begin{Highlighting}[]
\NormalTok{tbl }\OperatorTok{%>%}
\StringTok{  }\KeywordTok{group_by}\NormalTok{(id, classification) }\OperatorTok{%>%}
\StringTok{  }\KeywordTok{summarise}\NormalTok{(}\DataTypeTok{n =} \KeywordTok{n}\NormalTok{()) }\OperatorTok{%>%}
\StringTok{  }\KeywordTok{mutate}\NormalTok{(}\DataTypeTok{pct =}\NormalTok{ n }\OperatorTok{/}\StringTok{ }\KeywordTok{sum}\NormalTok{(n)) }\OperatorTok{%>%}
\StringTok{  }\KeywordTok{group_by}\NormalTok{(classification) }\OperatorTok{%>%}
\StringTok{  }\KeywordTok{summarise}\NormalTok{(}\StringTok{`}\DataTypeTok{Average percent}\StringTok{`}\NormalTok{ =}\StringTok{ }\KeywordTok{mean}\NormalTok{(pct) }\OperatorTok{*}\StringTok{ }\DecValTok{100}\NormalTok{) }\OperatorTok{%>%}
\StringTok{  }\KeywordTok{arrange}\NormalTok{(}\OperatorTok{-}\StringTok{`}\DataTypeTok{Average percent}\StringTok{`}\NormalTok{)}
\end{Highlighting}
\end{Shaded}

\begin{table}[t]

\caption{\label{tab:unnamed-chunk-30}Average percent of functions spent on each task.}
\centering
\begin{tabular}{lr}
\toprule
classification & Average percent\\
\midrule
data cleaning & 36.40\\
visualization & 23.17\\
exploratory & 21.32\\
setup & 18.87\\
modeling & 17.69\\
\addlinespace
import & 8.58\\
communication & 5.14\\
evaluation & 3.62\\
export & 0.82\\
\bottomrule
\end{tabular}
\end{table}

We can also examine most common functions in each classification.

\begin{Shaded}
\begin{Highlighting}[]
\NormalTok{func_counts <-}\StringTok{ }\NormalTok{tbl }\OperatorTok{%>%}
\StringTok{  }\KeywordTok{count}\NormalTok{(func, classification, }\DataTypeTok{sort =} \OtherTok{TRUE}\NormalTok{) }\OperatorTok{%>%}
\StringTok{  }\KeywordTok{ungroup}\NormalTok{()}

\NormalTok{func_counts}
\end{Highlighting}
\end{Shaded}

\begin{verbatim}
## # A tibble: 152 x 3
##    func      classification     n
##    <chr>     <chr>          <int>
##  1 summary   exploratory      361
##  2 lm        modeling         277
##  3 factor    data cleaning    141
##  4 select    data cleaning    138
##  5 library   setup            128
##  6 as.factor data cleaning    116
##  7 filter    data cleaning    107
##  8 aes       visualization     89
##  9 ggplot    visualization     82
## 10 lmer      modeling          80
## # ... with 142 more rows
\end{verbatim}

\begin{Shaded}
\begin{Highlighting}[]
\NormalTok{func_counts }\OperatorTok{%>%}
\StringTok{  }\KeywordTok{filter}\NormalTok{(classification }\OperatorTok{%in%}\StringTok{ }\KeywordTok{c}\NormalTok{(}\StringTok{"data cleaning"}\NormalTok{, }\StringTok{"exploratory"}\NormalTok{, }\StringTok{"modeling"}\NormalTok{, }\StringTok{"visualization"}\NormalTok{)) }\OperatorTok{%>%}
\StringTok{  }\KeywordTok{group_by}\NormalTok{(classification) }\OperatorTok{%>%}
\StringTok{  }\KeywordTok{top_n}\NormalTok{(}\DecValTok{5}\NormalTok{) }\OperatorTok{%>%}
\StringTok{  }\KeywordTok{ungroup}\NormalTok{() }\OperatorTok{%>%}
\StringTok{  }\KeywordTok{mutate}\NormalTok{(}\DataTypeTok{func =} \KeywordTok{reorder}\NormalTok{(func, n)) }\OperatorTok{%>%}
\StringTok{  }\KeywordTok{ggplot}\NormalTok{(}\KeywordTok{aes}\NormalTok{(func, n, }\DataTypeTok{fill =}\NormalTok{ classification)) }\OperatorTok{+}
\StringTok{  }\KeywordTok{theme_bw}\NormalTok{() }\OperatorTok{+}
\StringTok{  }\KeywordTok{geom_col}\NormalTok{(}\DataTypeTok{show.legend =} \OtherTok{FALSE}\NormalTok{) }\OperatorTok{+}
\StringTok{  }\KeywordTok{facet_wrap}\NormalTok{(}\OperatorTok{~}\NormalTok{classification, }\DataTypeTok{scales =} \StringTok{"free_y"}\NormalTok{) }\OperatorTok{+}
\StringTok{  }\KeywordTok{scale_x_discrete}\NormalTok{(}\KeywordTok{element_blank}\NormalTok{()) }\OperatorTok{+}
\StringTok{  }\KeywordTok{scale_y_continuous}\NormalTok{(}\StringTok{"Number of function calls in each classification"}\NormalTok{) }\OperatorTok{+}
\StringTok{  }\KeywordTok{coord_flip}\NormalTok{()}
\end{Highlighting}
\end{Shaded}

\begin{figure}
\centering
\includegraphics{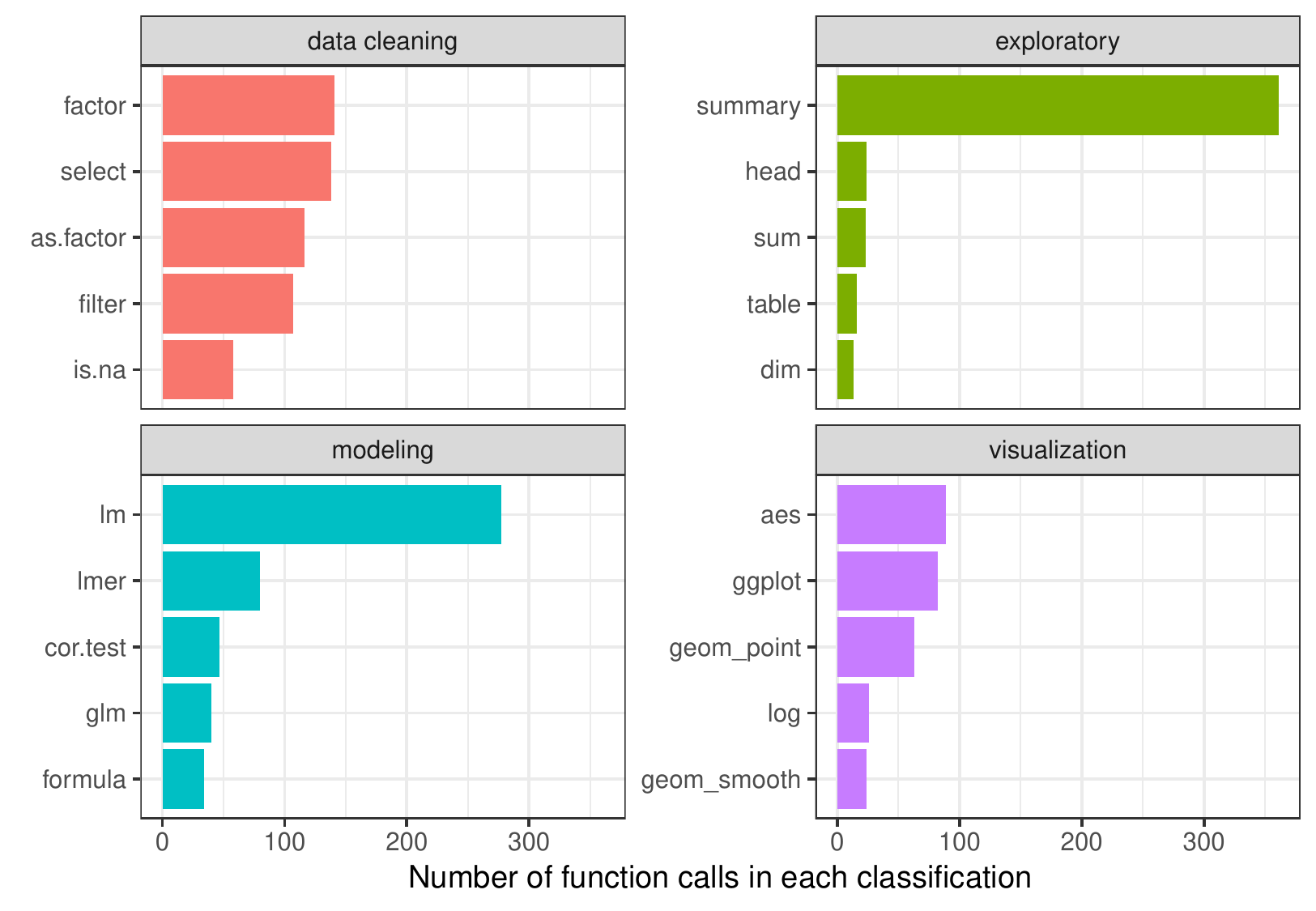}
\caption{Functions that contribute to data cleaning, exploratory
analysis, modeling and visualization classifications in p-hack-athon
trial}
\end{figure}

We could also examine a word cloud of the functions used, colored by the
classification. We can do this using the \textbf{wordcloud} library.

\begin{Shaded}
\begin{Highlighting}[]
\KeywordTok{library}\NormalTok{(wordcloud)}

\NormalTok{tbl }\OperatorTok{%>%}
\StringTok{  }\KeywordTok{count}\NormalTok{(func, classification) }\OperatorTok{%>%}
\StringTok{  }\KeywordTok{with}\NormalTok{(}
    \KeywordTok{wordcloud}\NormalTok{(func, n,}
      \DataTypeTok{colors =} \KeywordTok{brewer.pal}\NormalTok{(}\DecValTok{9}\NormalTok{, }\StringTok{"Set1"}\NormalTok{)[}\KeywordTok{factor}\NormalTok{(.}\OperatorTok{$}\NormalTok{classification)],}
      \DataTypeTok{random.order =} \OtherTok{FALSE}\NormalTok{,}
      \DataTypeTok{ordered.colors =} \OtherTok{TRUE}
\NormalTok{    )}
\NormalTok{  )}
\end{Highlighting}
\end{Shaded}

\begin{figure}
\centering
\includegraphics{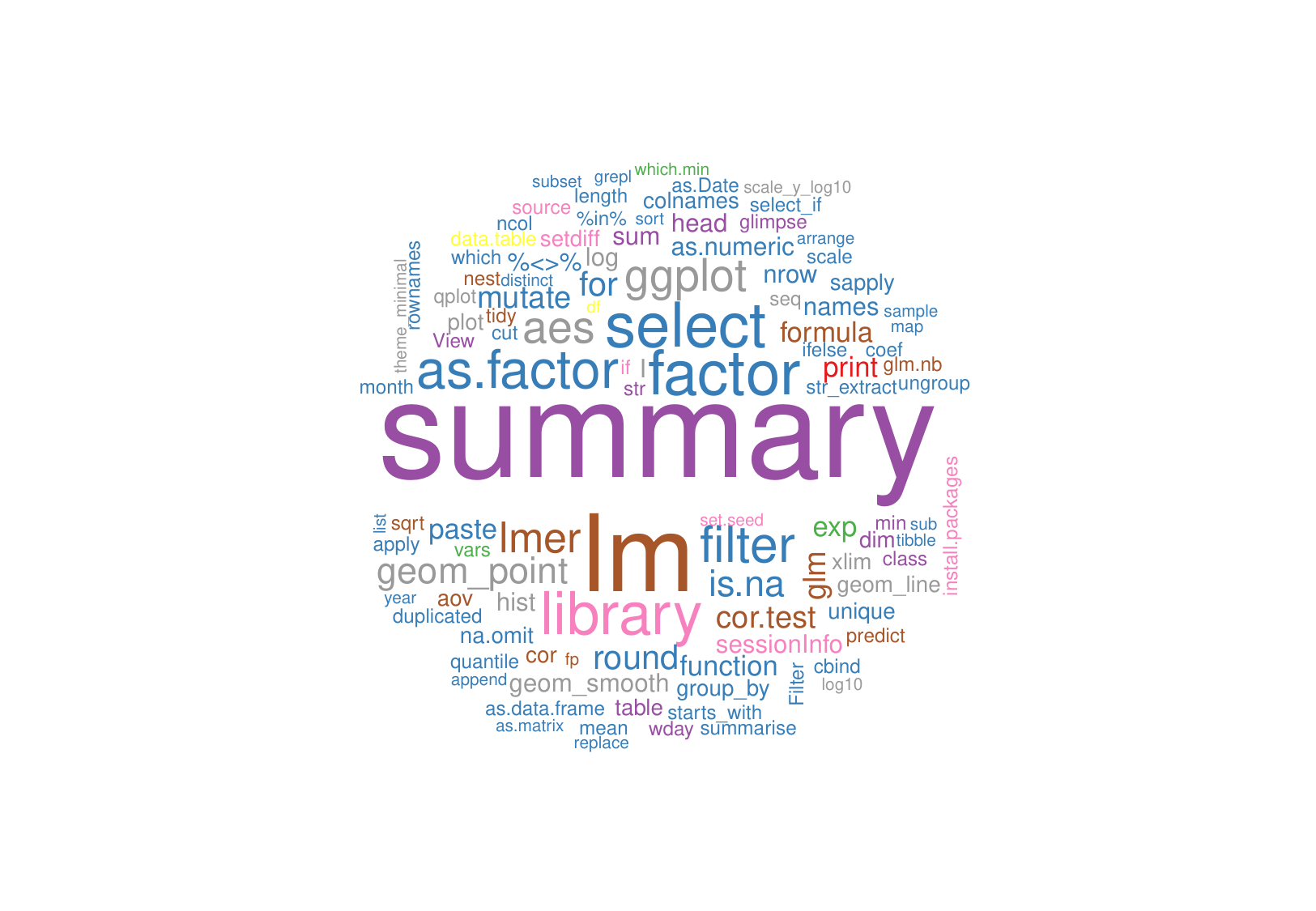}
\caption{Word cloud of functions used in the p-hack-athon trial, colored
by classification}
\end{figure}

\hypertarget{static-analysis}{%
\subsubsection{Static Analysis}\label{static-analysis}}

This second example demonstrates how to use the \textbf{matahari} and
\textbf{tidycode} packages to analyze data from a retrospective study,
or static R scripts. Here, we use the \texttt{read\_rfiles()} function
from the \textbf{tidycode} package. This wraps the
\texttt{dance\_recital()} \textbf{matahari} function and allows for
multiple file paths or urls to be read, resulting in a tidy data frame.
As an example, we are going to scrape all of the .R files from two of
the most widely used data manipulation packages, the \textbf{data.table}
package (Dowle and Srinivasan 2019) and the \textbf{dplyr} package. We
are going to use the \textbf{gh} package (Bryan and Wickham 2017) to
scrape these files from GitHub.

\hypertarget{setup-1}{%
\paragraph{Setup}\label{setup-1}}

We access the files via GitHub using the \texttt{gh()} function from the
\textbf{gh} package. This gives a list of download urls that can be
passed to the \texttt{read\_rfiles()} function from the
\textbf{tidycode} package.

\begin{Shaded}
\begin{Highlighting}[]
\KeywordTok{library}\NormalTok{(tidyverse)}
\KeywordTok{library}\NormalTok{(gh)}
\KeywordTok{library}\NormalTok{(tidycode)}

\NormalTok{dplyr_code <-}\StringTok{ }\KeywordTok{gh}\NormalTok{(}\StringTok{"/repos/tidyverse/dplyr/contents/R"}\NormalTok{) }\OperatorTok{%>%}
\StringTok{  }\NormalTok{purrr}\OperatorTok{::}\KeywordTok{map}\NormalTok{(}\StringTok{"download_url"}\NormalTok{) }\OperatorTok{%>%}
\StringTok{  }\KeywordTok{read_rfiles}\NormalTok{()}

\NormalTok{datatable_code <-}\StringTok{ }\KeywordTok{gh}\NormalTok{(}\StringTok{"/repos/Rdatatable/data.table/contents/R"}\NormalTok{) }\OperatorTok{%>%}
\StringTok{  }\NormalTok{purrr}\OperatorTok{::}\KeywordTok{map}\NormalTok{(}\StringTok{"download_url"}\NormalTok{) }\OperatorTok{%>%}
\StringTok{  }\KeywordTok{read_rfiles}\NormalTok{()}
\end{Highlighting}
\end{Shaded}

\hypertarget{data-cleaning}{%
\paragraph{Data Cleaning}\label{data-cleaning}}

We can combine these two tidy data frames. We will do some small data
manipulation, removing R calls that were either \texttt{NULL} or
\texttt{character}. For example, in the \textbf{dplyr} package some .R
files just reference data frames as a character string.

\begin{Shaded}
\begin{Highlighting}[]
\NormalTok{pkg_data <-}\StringTok{ }\KeywordTok{bind_rows}\NormalTok{(}
  \KeywordTok{list}\NormalTok{(}
    \DataTypeTok{dplyr =}\NormalTok{ dplyr_code,}
    \DataTypeTok{datatable =}\NormalTok{ datatable_code}
\NormalTok{  ),}
  \DataTypeTok{.id =} \StringTok{"pkg"}
\NormalTok{) }\OperatorTok{%>%}
\StringTok{  }\KeywordTok{filter}\NormalTok{(}
    \OperatorTok{!}\KeywordTok{map_lgl}\NormalTok{(expr, is.null),}
    \OperatorTok{!}\KeywordTok{map_lgl}\NormalTok{(expr, is.character)}
\NormalTok{  )}
\end{Highlighting}
\end{Shaded}

\hypertarget{analyze-r-functions}{%
\paragraph{Analyze R functions}\label{analyze-r-functions}}

Now we can use the \textbf{tidycode} \texttt{unnest\_calls()} function
to create a tidy data frame of the individual functions along with the
arguments used to create both packages. Notice here we are not
performing an anti join on ``stop functions''. For this analysis, we are
interested in examining some key differences in the commonly used
functions contained the two packages. Common operators may actually be
of interest, so we do not want to drop them from the data frame. We can
count the functions by package.

\begin{Shaded}
\begin{Highlighting}[]
\NormalTok{func_counts <-}\StringTok{ }\NormalTok{pkg_data }\OperatorTok{%>%}
\StringTok{  }\KeywordTok{unnest_calls}\NormalTok{(expr) }\OperatorTok{%>%}
\StringTok{  }\KeywordTok{count}\NormalTok{(pkg, func, }\DataTypeTok{sort =} \OtherTok{TRUE}\NormalTok{)}

\NormalTok{func_counts}
\end{Highlighting}
\end{Shaded}

\begin{verbatim}
## # A tibble: 1,163 x 3
##    pkg       func         n
##    <chr>     <chr>    <int>
##  1 datatable =         1640
##  2 dplyr     <-        1634
##  3 datatable if        1590
##  4 datatable {         1172
##  5 dplyr     {         1047
##  6 dplyr     function   724
##  7 datatable !          616
##  8 datatable <-         579
##  9 datatable [          564
## 10 datatable length     557
## # ... with 1,153 more rows
\end{verbatim}

Using this data frame, we can visualize which functions are most
commonly called in each package.

\begin{Shaded}
\begin{Highlighting}[]
\NormalTok{top_funcs <-}\StringTok{ }\NormalTok{func_counts }\OperatorTok{%>%}
\StringTok{  }\KeywordTok{group_by}\NormalTok{(pkg) }\OperatorTok{%>%}
\StringTok{  }\KeywordTok{top_n}\NormalTok{(}\DecValTok{10}\NormalTok{) }\OperatorTok{%>%}
\StringTok{  }\KeywordTok{ungroup}\NormalTok{() }\OperatorTok{%>%}
\StringTok{  }\KeywordTok{arrange}\NormalTok{(pkg, n) }\OperatorTok{%>%}
\StringTok{  }\KeywordTok{mutate}\NormalTok{(}\DataTypeTok{i =} \KeywordTok{row_number}\NormalTok{())}

\KeywordTok{ggplot}\NormalTok{(top_funcs, }\KeywordTok{aes}\NormalTok{(i, n, }\DataTypeTok{fill =}\NormalTok{ pkg)) }\OperatorTok{+}
\StringTok{  }\KeywordTok{theme_bw}\NormalTok{() }\OperatorTok{+}
\StringTok{  }\KeywordTok{geom_col}\NormalTok{(}\DataTypeTok{show.legend =} \OtherTok{FALSE}\NormalTok{) }\OperatorTok{+}
\StringTok{  }\KeywordTok{facet_wrap}\NormalTok{(}\OperatorTok{~}\NormalTok{pkg, }\DataTypeTok{scales =} \StringTok{"free"}\NormalTok{) }\OperatorTok{+}
\StringTok{  }\KeywordTok{scale_x_continuous}\NormalTok{(}
    \KeywordTok{element_blank}\NormalTok{(),}
    \DataTypeTok{breaks =}\NormalTok{ top_funcs}\OperatorTok{$}\NormalTok{i,}
    \DataTypeTok{labels =}\NormalTok{ top_funcs}\OperatorTok{$}\NormalTok{func,}
    \DataTypeTok{expand =} \KeywordTok{c}\NormalTok{(}\DecValTok{0}\NormalTok{, }\DecValTok{0}\NormalTok{)}
\NormalTok{  ) }\OperatorTok{+}
\StringTok{  }\KeywordTok{coord_flip}\NormalTok{()}
\end{Highlighting}
\end{Shaded}

\begin{figure}
\centering
\includegraphics{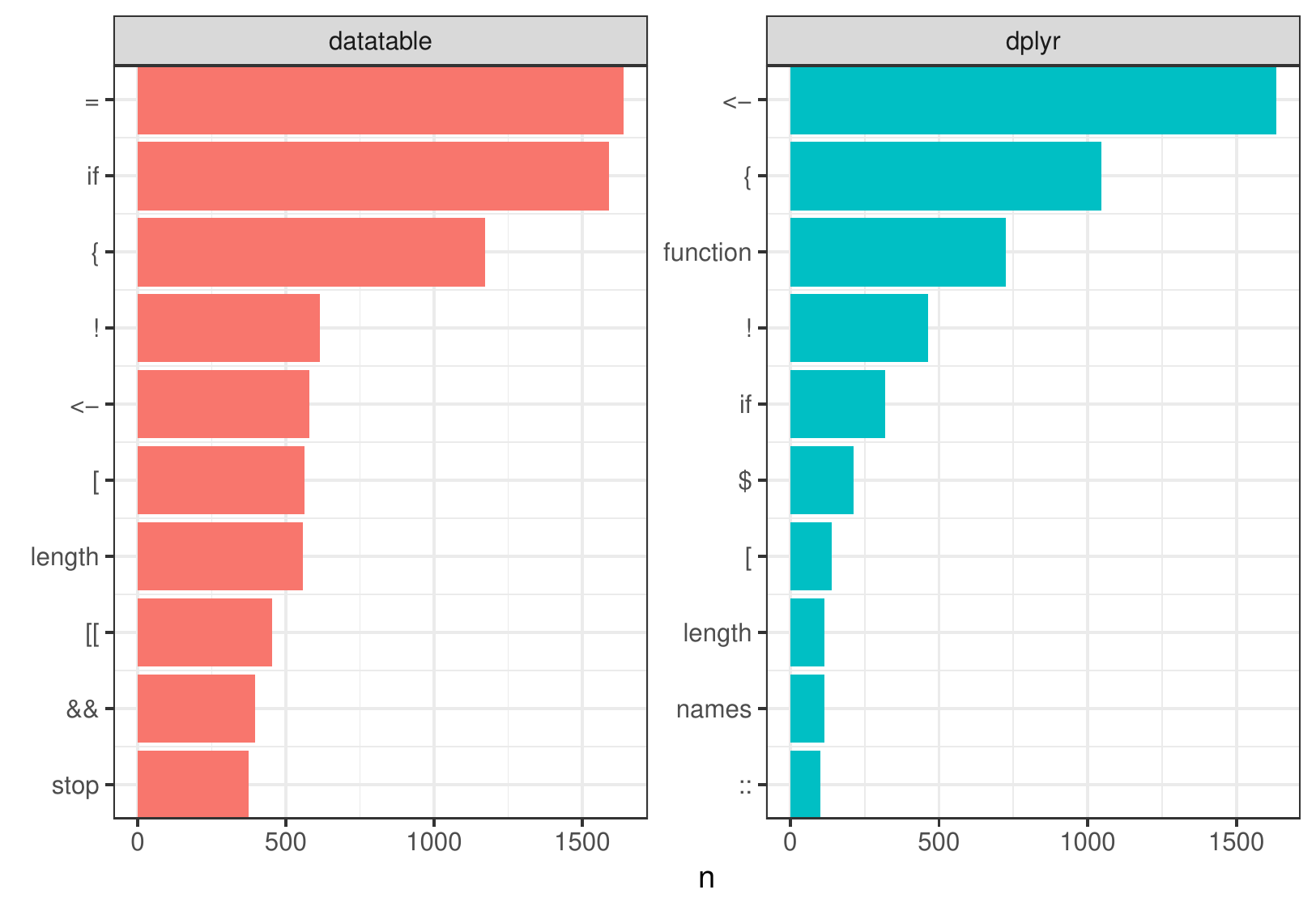}
\caption{Most frequent functions used in data.table and dplyr package
development.}
\end{figure}

We can glean a few interesting details from Figure 4. First, the
\textbf{data.table} authors sometimes use the \texttt{=} as an
assignment operator, resulting in this being the most frequent function
used. The \textbf{dplyr} authors always use \texttt{\textless{}-} for
assignment, therefore this is the most frequent function seen in this
package (Wickham 2019). Additionally, the \textbf{dplyr} authors often
create modular code as a combination of small functions to complete
specific tasks. This may explain why \texttt{function} is the third most
frequent R call in this package, and less prevalent in the
\textbf{data.table} package. This just serves as a glimpse of what can
be accomplished with these tools.

\hypertarget{discussion}{%
\subsection{Discussion}\label{discussion}}

We have designed a framework to analyze the data analysis pipeline and
created two R packages that allow for the study of data analysis code
conducted in R. We present two packages, \textbf{matahari}, a package
for logging everything that is typed in the R console or in an R script,
and \textbf{tidycode}, a package with tools to allow for analyzing R
calls in a tidy manner. These tools can be applied both to prospective
studies, where a researcher can intentionally record code typed by
participants, and retrospectively, where the researcher can
retrospectively analyze code. We believe that these tools will help
shape the next phase of reproducibility and replicability, allowing the
analysis of code to inform data science pedagogy, examine how
collaborates conduct data analyses, and explore how current software
tools are being utilized.

\hypertarget{acknowledgements}{%
\subsection{Acknowledgements}\label{acknowledgements}}

We would like to extend a special thank you to the members of the Leek
Lab at Johns Hopkins Bloomberg School of Public Health as well as
volunteers who used the ``classify'' shiny application for helping
classify R functions.

\hypertarget{references}{%
\subsection*{References}\label{references}}
\addcontentsline{toc}{subsection}{References}

\hypertarget{refs}{}
\leavevmode\hypertarget{ref-gh}{}%
Bryan, Jennifer, and Hadley Wickham. 2017. \emph{Gh: 'GitHub' 'Api'}.
\url{https://CRAN.R-project.org/package=gh}.

\leavevmode\hypertarget{ref-buja2009statistical}{}%
Buja, Andreas, Dianne Cook, Heike Hofmann, Michael Lawrence, Eun-Kyung
Lee, Deborah F Swayne, and Hadley Wickham. 2009. ``Statistical Inference
for Exploratory Data Analysis and Model Diagnostics.''
\emph{Philosophical Transactions of the Royal Society A: Mathematical,
Physical and Engineering Sciences} 367 (1906). The Royal Society
Publishing: 4361--83.

\leavevmode\hypertarget{ref-datatable}{}%
Dowle, Matt, and Arun Srinivasan. 2019. \emph{Data.table: Extension of
`Data.frame`}. \url{https://CRAN.R-project.org/package=data.table}.

\leavevmode\hypertarget{ref-Goecks:2010ea}{}%
Goecks, Jeremy, Anton Nekrutenko, James Taylor, and Galaxy Team. 2010.
``Galaxy: a comprehensive approach for supporting accessible,
reproducible, and transparent computational research in the life
sciences.'' \emph{Genome Biology} 11 (8).

\leavevmode\hypertarget{ref-Ioannidis:2014cm}{}%
Ioannidis, John P A, Marcus R Munafo, Paolo Fusar-Poli, Brian A Nosek,
and Sean P David. 2014. ``Publication and other reporting biases in
cognitive sciences: detection, prevalence, and prevention.''
\emph{Trends in Cognitive Sciences} 18 (5): 235--41.

\leavevmode\hypertarget{ref-Leek:2015kp}{}%
Leek, Jeffrey T, and Roger D Peng. 2015. ``Opinion: Reproducible
research can still be wrong: Adopting a prevention approach.''
\emph{Proceedings of the National Academy of Sciences} 112 (6): 1645--6.

\leavevmode\hypertarget{ref-loy2016variations}{}%
Loy, Adam, Lendie Follett, and Heike Hofmann. 2016. ``Variations of Q--Q
Plots: The Power of Our Eyes!'' \emph{The American Statistician} 70 (2).
Taylor \& Francis: 202--14.

\leavevmode\hypertarget{ref-loy2017model}{}%
Loy, Adam, Heike Hofmann, and Dianne Cook. 2017. ``Model Choice and
Diagnostics for Linear Mixed-Effects Models Using Statistics on Street
Corners.'' \emph{Journal of Computational and Graphical Statistics} 26
(3). Taylor \& Francis: 478--92.

\leavevmode\hypertarget{ref-majumder2013validation}{}%
Majumder, Mahbubul, Heike Hofmann, and Dianne Cook. 2013. ``Validation
of Visual Statistical Inference, Applied to Linear Models.''
\emph{Journal of the American Statistical Association} 108 (503). Taylor
\& Francis Group: 942--56.

\leavevmode\hypertarget{ref-McNutt:2014bq}{}%
McNutt, M. 2014. ``Reproducibility.'' \emph{Science} 343 (6168):
229--29.

\leavevmode\hypertarget{ref-Miguel:2014hr}{}%
Miguel, E, C Camerer, K Casey, J Cohen, K M Esterling, A Gerber, R
Glennerster, et al. 2014. ``Promoting Transparency in Social Science
Research.'' \emph{Science} 343 (6166): 30--31.

\leavevmode\hypertarget{ref-Nosek:2015bz}{}%
Nosek, B A, G Alter, G C Banks, D Borsboom, S D Bowman, S J Breckler, S
Buck, et al. 2015. ``Promoting an open research culture.''
\emph{Science} 348 (6242): 1422--5.

\leavevmode\hypertarget{ref-Peng:2011et}{}%
Peng, Roger D. 2011. ``Reproducible Research in Computational Science.''
\emph{Science} 334 (6060): 1226--7.

\leavevmode\hypertarget{ref-Richard:2014cn}{}%
Richard, Blaustein. 2014. ``Reproducibility Undergoes Scrutiny.''
\emph{BioScience} 64 (4): 368--68.

\leavevmode\hypertarget{ref-Sidi:2018hk}{}%
Sidi, Yulia, and Ofer Harel. 2018. ``The treatment of incomplete data:
Reporting, analysis, reproducibility, and replicability.'' \emph{Social
Science \& Medicine} 209 (July): 169--73.

\leavevmode\hypertarget{ref-silberzhan2018many}{}%
Silberzhan, Raphael, Eric L Uhlmann, Daniel P Martin, Pasquale Anselmi,
Frederick Aust, Eli Awtrey, Štěpán Bahník, et al. 2018. ``Many Analysts,
One Data Set: Making Transparent How Variations in Analytic Choices
Affect Results.'' \emph{Advances in Methods and Practices in
Psychological Science} 1 (3). Sage Publications Sage CA: Los Angeles,
CA: 337--56.

\leavevmode\hypertarget{ref-silge2016tidytext}{}%
Silge, Julia, and David Robinson. 2016. ``Tidytext: Text Mining and
Analysis Using Tidy Data Principles in R.'' \emph{The Journal of Open
Source Software} 1 (3): 37.

\leavevmode\hypertarget{ref-silge2017text}{}%
---------. 2017. \emph{Text Mining with R: A Tidy Approach}. " O'Reilly
Media, Inc.".

\leavevmode\hypertarget{ref-Waltemath:2016jw}{}%
Waltemath, Dagmar, and Olaf Wolkenhauer. 2016. ``How Modeling Standards,
Software, and Initiatives Support Reproducibility in Systems Biology and
Systems Medicine.'' \emph{Ieee Transactions on Biomedical Engineering}
63 (10): 1999--2006.

\leavevmode\hypertarget{ref-wickham2015r}{}%
Wickham, Hadley. 2015. \emph{R Packages: Organize, Test, Document, and
Share Your Code}. " O'Reilly Media, Inc.".

\leavevmode\hypertarget{ref-tidyverse}{}%
---------. 2017. \emph{Tidyverse: Easily Install and Load the
'Tidyverse'}. \url{https://CRAN.R-project.org/package=tidyverse}.

\leavevmode\hypertarget{ref-tidystyle}{}%
---------. 2019. \emph{The Tidyverse Style Guide}.
\url{https://style.tidyverse.org}.

\leavevmode\hypertarget{ref-wickham2015visualizing}{}%
Wickham, Hadley, Dianne Cook, and Heike Hofmann. 2015. ``Visualizing
Statistical Models: Removing the Blindfold.'' \emph{Statistical Analysis
and Data Mining: The ASA Data Science Journal} 8 (4). Wiley Online
Library: 203--25.

\leavevmode\hypertarget{ref-wickham2016r}{}%
Wickham, Hadley, and Garrett Grolemund. 2016. \emph{R for Data Science:
Import, Tidy, Transform, Visualize, and Model Data}. " O'Reilly Media,
Inc.".

\end{document}